\documentclass[intlimits, twoside, a4paper]{article}

\usepackage[cp1251]{inputenc}

\usepackage[eqsecnum]{cmpj3}

\issue{2023}{26}{4}{43604}
\doinumber{10.5488/CMP.26.43604}

\title[Nonlinear background corrections to dielectric permittivity of ferroics and multiferroics]
{Nonlinear background corrections to dielectric permittivity of ferroics and multiferroics}

\author[I. S. Girnyk, B. I. Horon, V. B. Kapustianyk, O. S. Kushnir, R. Y. Shopa]{I.~S.~Girnyk\refaddr{label1}, B.~I.~Horon\orcid{0000-0002-4595-5789}\refaddr{label1, label2}\thanks{Corresponding Author: \email{bohdan.horon@lnu.edu.ua}}, V.~B.~Kapustianyk\orcid{0000-0001-7830-5670}\refaddr{label1}, O.~S.~Kushnir\orcid{0000-0002-1545-7666}\refaddr{label2}, R.~Y.~Shopa\orcid{0000-0002-1089-5050}\refaddr{label4}}
\addresses{
\addr{label1} Physics Department, Ivan Franko National University of Lviv, 50 Drahomanov Street, 79005 Lviv, Ukraine
\addr{label2} Electronics and Computer Technologies Department, Ivan Franko National University of Lviv, 107 Tarnavskyi Street, 79017 Lviv, Ukraine
\addr{label4} Department of Complex Systems, National Centre for Nuclear Research, 05-400 Otwock-\'{S}wierk, Poland}

\date{Received March 27, 2023, in final form August 16, 2023}
\begin{document}

\maketitle

\begin{abstract}
    Temperature measurements of dielectric permittivity are performed for nonstoichiometric ferroelectric lead germanate Pb$_{4.95}$Ge$_3$O$_{11}$ and multiferroic solid solution [N(C$_2$H$_5$)$_4$]$_2$CoClBr$_3$.
    Unlike the heat capacity data, the analysis of the dielectric permittivity of ferroics is usually performed at the assumption that the dielectric `background' is negligible  compared with its critical part.
    In this work we quantitatively interpret  the dielectric properties of the single crystals mentioned above and the appropriate literature data for multiferroic Sr$_2$IrO$_4$ crystals, using generalized Curie-Weiss formulas that combine corrections due to a nonlinear temperature-dependent dielectric background, a modified critical index of electric susceptibility, and a diffuse character of phase transition.
    We argue that taking  account of the temperature dependent dielectric background can improve notably the quantitative analysis of PTs for a number of classes of the ferroic materials.
\printkeywords
\end{abstract}

\section{Introduction} \label{sec:intro}

Temperature behavior of different physical characteristics, in particular dielectric ones, around the points of phase transitions (PTs) in ferroics and multiferroics is an extensively explored problem of condensed matter physics~\cite{linesglass1977, wangabliuabren2009}.
The temperature dependence of dielectric permittivity  can be influenced by thermal fluctuations, long-range dipolar correlations, inhomogeneity of a solid due to defects, its structural disorder and `diffuseness' of the PT, contributions of domain walls, etc.
Distinguishing among all those factors is not a simple task, particularly in a complicated case of multiferroics, for which there are some indications that the critical indices can be modified due to magnetoelectric coupling~\cite{bahooshwesselinowa2012}, diffuseness of PTs and a relaxor-like behavior\cite{ramondetal2005, chikaraetal2009, mukherjeeetal2010}.

In general, quantitative analyses of critical behavior of the dielectric permittivity with rigorous statistically based techniques are rarer  compared with those known for the heat capacity (see, e. g.,~\cite{ahlerskornblit1975}).
In many works only qualitative features of the  dependence are estimated (see, e. g.,~\cite{chikaraetal2009}), with no detailed and statistically grounded derivation of critical indices and amplitudes for the symmetric and asymmetric phases.
We believe that, besides the objective factors mentioned above, the other reason can be the so-called dielectric `background', which is not associated with the PT.
Its occurrence and some relevant mechanisms were realized long ago~\cite{linesglass1977, rupprechtbell1964}.
In particular, it would be natural to expect that the relative importance of the dielectric background should not be completely neglected for a large number of substances.
These are weakly polar ferroics~\cite{tagantsev1987, sandvoldcourtens1983}, ferroelectric-dielectric composites~\cite{shermanetal2006}, finite-sized or confined~\cite{morozovskaetal2010, tagantsevgerrasetter2008} systems such as thin-film or nanoscale ferroelectrics~\cite{bratkovskylevanyuk2009, zhengwoo2009, zhengwoo2009_2}, improper ferroelectrics~\cite{cmp2022} and ferroeleastics~\cite{ferro1993} in which the dielectric anomaly has a secondary character, ferroelectrics with relatively high defect concentrations~\cite{pts2007, ujpo2008} or nonstoichiometry~\cite{girnykklymovychkushnirshopa2014}, and mutiferroics~\cite{foxtilleyscottguggenheim1980, nenertetal2008, schrettleetal2008}.
For very different reasons, the critical part $\varepsilon_{\text{cr}}(T)$ can be small enough in these ferroics, thus imposing a greater relative contribution of the background.

The arguments mentioned above justify a further elaboration of the PT-independent contribution to the dielectric function.
Unfortunately, the researchers in the field often neglect $\varepsilon_B$ as a term small against $\varepsilon_{\text{cr}}$ or, at the most, use the simplest approximation $\varepsilon_B = \mathrm{const}$ (see, e. g.,~\cite{shermanetal2006, morozovskaetal2010, zhengwoo2009_2}), although situations can happen when the background is temperature-dependent or even nonlinear in temperature.
Then, common linear-regression or graphical techniques for deriving crucial PT parameters (see, e.g.,~\cite{ramondetal2005, morrisonsinclairwest1999, kovalalemanybriancinbrunckova2003, correakumarkatiyar2007}) cannot be employed, giving way to much more complex nonlinear fitting~\cite{ahlerskornblit1975}.
However, the latter situation is still rare.
Unlike the studies of the critical PT index $\alpha$ of the heat capacity, the critical index $\gamma$ associated with the dielectric permittivity is usually derived such that the dielectric background is disregarded or considered to be a constant, and the researchers examine only the critical part of the  dependence.

In this work we report experimental studies and phenomenological interpretation of three different $\varepsilon(T)$ functions, with the purpose of combining the possible effects of nonlinear background, non-unit critical index of susceptibility, and diffuse PT.
As examples, we have chosen single crystals of a ferroic lead germanate (PGO) crystal with nonstoichiometry and a multiferroic tetraethylammonium tetrahalogenometallic compound [N(C$_2$H$_5$)$_4$]$_2$CoClBr$_3$ (TEACCB-3).
Finally, the resources of our approach are briefly illustrated on the  dependence for multiferroic Sr$_2$IrO$_4$ crystals, which is taken from the work~\cite{chikaraetal2009}.

\section{Materials, methods and results} \label{sec:materials-and-methods}

According to the PbO and GeO$_2$ contents in charge, nonstoichiometric PGO can be described using a conditional formula Pb$_{4.95}$Ge$_3$O$_{11}$.
This PGO compound represents a ferroelectric with a second-order PT at the Curie point $T_{\textrm{C}} \approx 435$~K (the symmetry $P6 \leftrightarrow P3$), in which Pb vacancies of a preset concentration were created~\cite{ermakovduda2010, girnykklymovychkushnirshopa2014}.
Strong dipole-dipole correlations must have been available in this uniaxial ferroelectric.
The solid solution of TEACCB-3 which is known to be a magnetic multiferroic belongs to A$_2$BX$_4$ family of compounds, where A is an organic cation, B is a metal, and X is a halogen~\cite{izumi1990, kundysetal2010, kapustianyk2013}.
It reveals a second-order ferroelectric PT of an order-disorder type at $T_{\textrm{C}}\approx257$~K from the room-temperature phase $P4_2/\text{nmc}$, and there are indications to a magnetodielectric shift in the vicinity of this PT~\cite{kundysetal2010, kapustianyk2013}.
According to the literature, PGO and TEACCB-3 reveal no clear traces of a first-order character of the PTs, so that we do not distinguish between their Curie ($T_{\textrm{C}}$) and Curie-Weiss ($T_0$) temperature points.

Samples for the dielectric measurements were cut in the shape of parallelepipeds with the sizes $\sim 5 \times 5 \times 1$~mm$^3$.
The real part $\varepsilon$  of the low-field dielectric permittivity was measured using an automated capacitive apparatus (the operating voltage $\sim 1$~V applied along the polar $c$ axis; the frequencies $f=500$~kHz for PGO and $f = 100$~kHz for TEACCB-3).
In both cases, the frequency of the measuring electric field was chosen in the range where the contributions of fundamental dielectric dispersion and the dielectric dispersion caused by domain-wall dynamics were expected to be negligible.
A similar approach was used for phenomenological description of the magnetoelectric effect in another ferroelectric with alkylammonium cation, NH$_2$(CH$_3$)$_2$Al$_{1-x}$Cr$_x$(SO$_4$)$_2\times6$H$_2$O~\cite{kapustianyketal2019}.
The dielectric measurements were performed on mechanically free crystals.
The samples were heated at the rate $\rd T/\rd t =$~10--50~K/h and the temperature tolerance was equal to $\sim 0.1$~K. 

\begin{figure}[htb]
    \centering
    \includegraphics[width=0.7\textwidth]{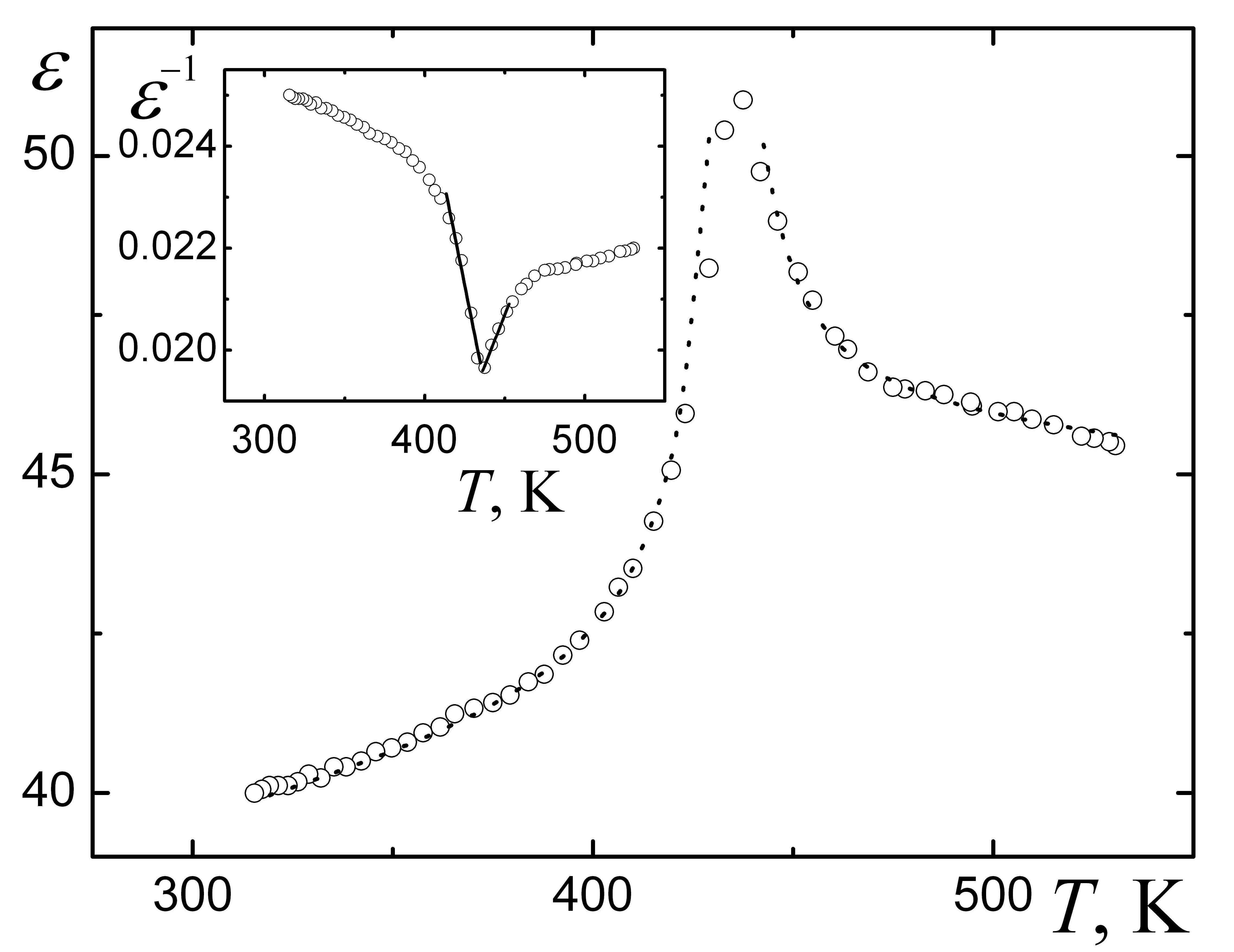}
    \caption{Dependence $\varepsilon(T)$ for the PGO crystals (see the text).
    Insert: $\varepsilon^{-1}(T)$ dependence.
    Straight lines correspond to the Cuire-Weiss law near $T_{\textrm{C}}$ and dot curves correspond to formula~(\ref{eq:eps-gen-cw}) with $\gamma=1$.}
    \label{fig:eps-PGO}
\end{figure}

Figure~\ref{fig:eps-PGO} shows the temperature dependence of the dielectric permittivity for the PGO crystals. The nonstoichiometry broadens and drastically suppresses  the anomaly at the PT (see also~\cite{ermakovduda2010, girnykklymovychkushnirshopa2014}).
Nonetheless, diffuseness of the PT is not clearly seen.
The simplest Curie-Weiss law ($\varepsilon = C / (T - T_{\textrm{C}})$, with a constant $C$) is seemingly fulfilled only in the closest vicinity of the PT (in the region $10^{-2} < t < 4 \cdot 10^{-2}$  of reduced temperatures $t = (T - T_{\textrm{C}})/T_{\textrm{C}}$ --- see figure~\ref{fig:eps-PGO}, insert), although the minimum $\varepsilon^{-1}$ values are too large.
At the same time, the ratio of slopes for the ferroelectric and paraelectric phases ($C_{-}/C_{+} \approx 1.97$) is very close to the Landau-theory result of two.
We suppose that `violation' of the Curie-Weiss law in a wider temperature range (at least at $t < 10^{-1}$) is only a seeming effect due to the neglect of a relatively important temperature-dependent dielectric background.

\begin{figure}[htb]
    \centering
    \includegraphics[width=0.7\textwidth]{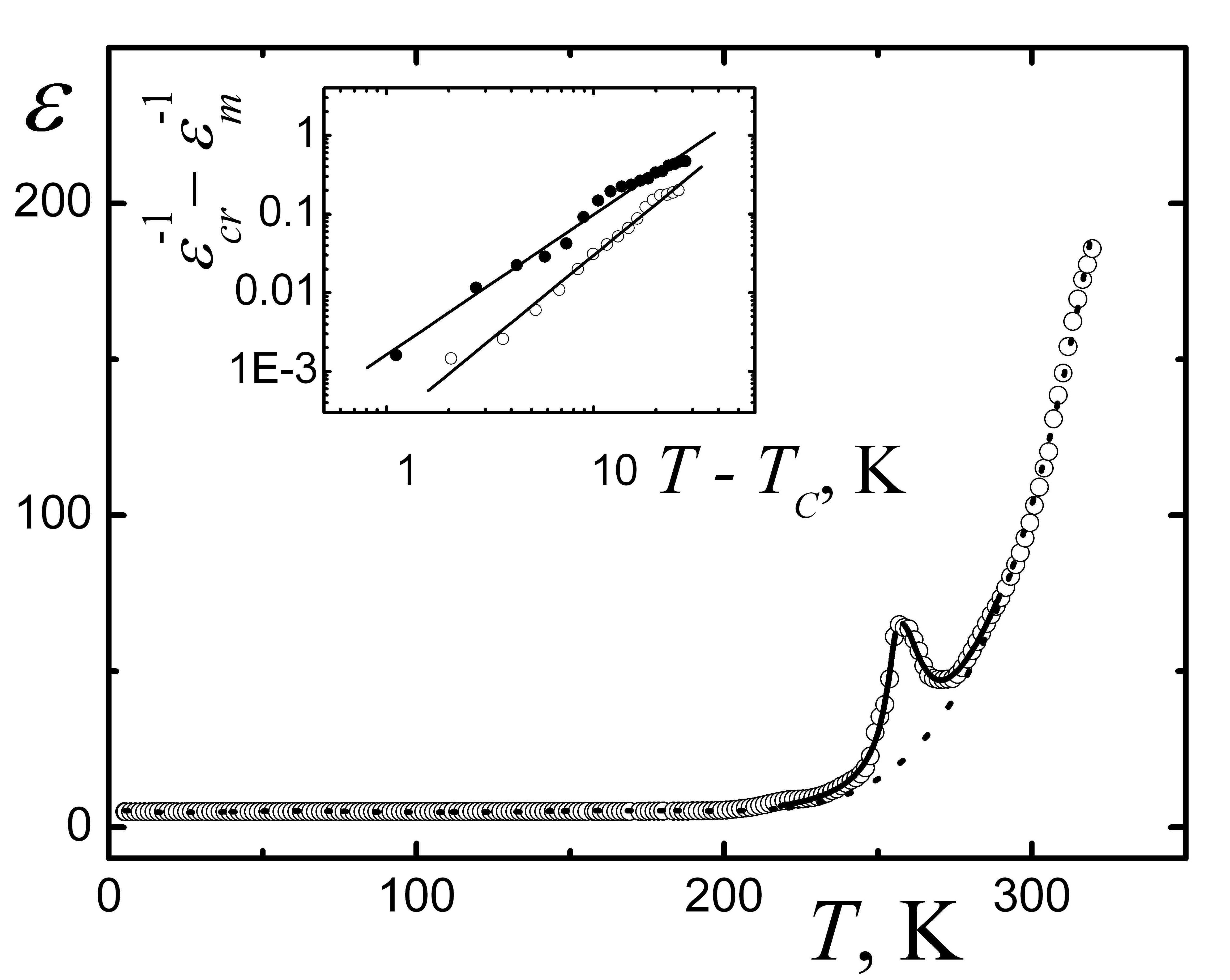}
    \caption{Dependence $\varepsilon(T)$ for the TEACCB-3 crystals (see the text).
    Not all of the data points are displayed.
    Dot curve corresponds to $\varepsilon_B(T)$ and solid curves correspond to formula~(\ref{eq:eps-dif-pt}).
    Insert: log-log dependences $1/\varepsilon_{\text{cr}} - 1/\varepsilon_\text{m}$ vs. $(T - T_{\textrm{C}})$ for $T < T_{\textrm{C}}$ ($\bullet$) and $T > T_{\textrm{C}}$ ($\circ$).}
    \label{fig:eps-TEACCB}
\end{figure}

The $\varepsilon(T)$ curve for the TEACCB-3 crystals is shown in figure~\ref{fig:eps-TEACCB}.
The dielectric peak at the PT is notably diffused and the background $\varepsilon_B$ evidently dominates at $T > 285$~K.
Deviations of the reciprocals $1/\varepsilon(T)$ calculated with figure~\ref{fig:eps-TEACCB} from the Curie-Weiss law are very serious. 

\section{Data interpretation and discussion} \label{sec:data-interpretation}

A general trend to increasing permittivity with increased temperature seen from figure~\ref{fig:eps-PGO} and figure~\ref{fig:eps-TEACCB} has nothing to do with the PT.
For PGO, this is caused most likely, by space-charge effects and off-center structural substitutions, being accompanied by growing dielectric losses and conductivity (see~\cite{girnykklymovychkushnirshopa2014}).
Probably, $\varepsilon_B$ in TEACCB-3 is related to ionic or, maybe, proton conductivity (see~\cite{kundysetal2010, kapustianyk2013}).
Under the condition of damped-down dielectric maxima $\varepsilon_{\text{max}}$ (e. g., we have $\varepsilon_{\text{max}} \approx 51$ for PGO), which  originate from the defects due to nonstoichiometry in PGO or structural disorder in TEACCB-3. This background contribution requires a proper consideration.
In the case of PGO, we assume the simplest nonlinear background, $\varepsilon_B(T) = \varepsilon_0 + \varepsilon_1 T + \varepsilon_2 T^2$, with temperature-independent $\varepsilon_i$'s.
This yields in the relation
\begin{equation} \label{eq:eps-gen-cw}
    \varepsilon(T) = \varepsilon_B(T) + C/(T - T_{\textrm{C}})^{\gamma},
\end{equation}
where retaining  the most general case $\gamma \neq 1$ for the critical index $\gamma$ corresponds to modification of the Curie-Weiss law driven by the structural defects or some other factors, which cannot be excluded.
At least, it is known that crystal imperfections can considerably alter the critical index $\alpha$  of the heat capacity~\cite{levanyukosipovsigovsobyanin1979, scotthabbalhidaka1982}.

On the other hand, following  the assumption of Gaussian spatial distribution of the Curie temperatures in the case of diffuse PTs, one can arrive at the approximate expression $1/\varepsilon_{\text{cr}} - 1/\varepsilon_\text{m} = (T - T_{\textrm{C}})^{\gamma}/C^{\prime}$ for the critical part $\varepsilon_{\text{cr}}(T)$ of the dielectric function~\cite{morrisonsinclairwest1999, smolenskii1970, uchinonomura1982, pilgrimsutherlandwinzer1990}, where $\varepsilon_\text{m}$ is a maximum of the critical part of the dielectric permittivity at $T_{\textrm{C}}$, the `diffuseness index' $\gamma$ varies from 1 for the ferroelectric PT with no diffuseness to 2 for the relaxor-type PTs, and $C^{\prime}$ is a constant.
Note that $\varepsilon_\text{m}$ gives a maximum of the dielectric permittivity only when there is no background $\varepsilon_B$.
Taking the term $\varepsilon_B(T)$ into consideration, one gets the relation
\begin{equation} \label{eq:eps-dif-pt}
    \varepsilon(T) = \varepsilon_B(T) + \frac{\varepsilon_\text{m}}{1 + (T - T_{\textrm{C}})^{\gamma}/(2\delta^2)},
\end{equation}
where $\delta$ characterizes a `broadening' of dielectric anomaly ($C^{\prime} = 2\varepsilon_\text{m}\delta^2$).
Its meaning becomes the most transparent in the case of a `Lorentzian-like' limit for formula~(\ref{eq:eps-dif-pt}) (i. e., under condition $\gamma=2$).
Finally, we note that the relation $\varepsilon_{\text{max}} = \varepsilon_B (T_{\textrm{C}}) + \varepsilon_\text{m}$ holds true in the case of a nonzero dielectric background.
The approximation given by formula~(\ref{eq:eps-dif-pt}) is accurate enough close to the $T_{\textrm{C}}$ point ($\varepsilon > (2/3) \varepsilon_\text{m}$~\cite{pilgrimsutherlandwinzer1990}).
Notice that the limit of non-diffuse PTs with the Curie-Weiss law can be recovered via $\gamma \rightarrow 1$, $\delta \rightarrow 0$, $\varepsilon_\text{m} \rightarrow \infty$ and $\varepsilon_\text{m}\delta^2 \rightarrow \mathrm{const}$.

Nonlinear least-squares fitting of our data for PGO by formula~(\ref{eq:eps-dif-pt}) with the quadratic $\varepsilon_B(T)$ function has revealed a tendency given by the approximate equality $\gamma \sim 1$, with physically unrealistic value $\varepsilon_0 < 0$ (see also the discussion in the work~\cite{rupprechtbell1964}).
Since the hypothesis of $\gamma \neq 1$ has not been proved in terms of statistics, we  further concentrated on formula~(\ref{eq:eps-gen-cw}) with $\gamma = 1$ (see dot lines in figure~\ref{fig:eps-PGO} and the data presented in table~\ref{tab:fit-results}).
The corresponding standard deviation values obtained by us are close to the experimental accuracy and the goodness-of-fit parameters $R^2$ are high enough for both the ferroelectric and paraelectric phases.
Close $\varepsilon_B$ values obtained for $T < T_{\textrm{C}}$ and $T > T_{\textrm{C}}$, and a $\sim 50$\% higher ferroelectric-phase $C$ value are also reasonable.
The same refers to the PT points $T_{\textrm{C}}$ derived in the independent fits for the high- and low-temperature phases, which are less than 1\%~different.
Note also that there is no solid evidence for treating this difference as a sign of a first-order character of the PT, i.e., a measure of a difference between the Curie and Curie-Weiss temperatures.
As already mentioned above, no clear indication of this effect can be found in the literature for all of the crystals under study.
Just as with the differences between the fitting parameters $\varepsilon_i$ for the paraelectric and ferroelectric phases, here we most probably deal with some experimental and/or fitting errors.
At least partly, the same reasons explain the fact $C_{-} / C_{+} \neq 2$ (see table~\ref{tab:fit-results}).

\begin{table}[htb]
    \caption{Parameters of fitting of $\varepsilon(T)$ dependence for the PGO crystals using formula~(\ref{eq:eps-gen-cw}) with $\gamma = 1$.}
    \label{tab:fit-results}
    \vspace{3mm}
    \centering
    \begin{tabular}{|c|c|c|}
    \hline
    Parameter & Ferroelectric phase & Paraelectric phase \\
    \hline
    $\varepsilon_0$ & 33.6 & 39.4 \\
    $\varepsilon_1$, K$^{-1}$ & 0.021 & 0.028 \\
    $\varepsilon_2$, K$^{-2}$ & $-1\cdot 10^{-5}$ & $-3 \cdot 10^{-5}$ \\
    Curie-Weiss constant $C$, K & 81.0 & 54.2 \\
    Curie temperature $T_{\textrm{C}}$, K & 437.6 & 433.6 \\
    Mean-square deviation $SD$ & 0.05 & 0.09 \\
    Coefficient of determination $R^2$ & 0.998 & 0.993 \\
    \hline
    \end{tabular}
\end{table}

Notice that only the fitting parameters consistent for both structural phases, as is our case, can be regarded as physically meaningful in the analysis of critical behavior, but not the data obtained for a single phase only (see~\cite{ahlerskornblit1975}).
Some deviations from the theory are observed only in the closest vicinity of the PT (at $|t| < 2 \cdot 10^{-2}$) where there are a few data points.
Hence, the disagreement of the results for PGO with the classical theory, at least those referring to the interval $4 \cdot 10^{-2} < |t| < 10^{-1}$ (see insert in figure~\ref{fig:eps-PGO}), are a pure consequence of improper disregard of the dielectric background.
As a result, the dielectric data can be successfully explained in terms of the Curie-Weiss behavior with a non-negligible temperature-dependent background, rather than by assuming a diffuse PT.
Note that the above result somewhat differs from the conclusion drawn in our earlier work~\cite{elit2018} for the $\varepsilon(T)$ function obtained at a different electric-field frequency.
However, only the particular cases of formula~(\ref{eq:eps-gen-cw}) with $\varepsilon_B = 0$ and formula~(\ref{eq:eps-dif-pt}) with $\varepsilon_B = \mathrm{const}$ were compared in the study~\cite{elit2018} as theoretical models.

We took the advantage of a wide temperature range measured for TEACCB-3 and calculated the dielectric background, using a high-order polynomial and having excluded the data points from the region of $\sim 70$~K around the PT.
This has freed us from a necessity to employ complicated multi-parametric nonlinear fitting techniques.
Notice that we  also tried a general formula $\varepsilon_B(T) = \varepsilon_0 + C_0 [\exp{(\hbar\omega_0/kT)} - 1]^{-1}$~\cite{foxtilleyscottguggenheim1980}, with constant $\varepsilon_0$ and $C_0$, and the frequency $\omega_0$ associated with a soft optical branch. However, large negative $\varepsilon_0$ and unrealistically high $\omega_0$ ($\sim 2200$~cm$^{-1}$) obtained by us testify that this formula does not fit the experimental data.
This is not surprising since, being derived in the same approximation as a quantum generalization of classical Curie-Weiss criticality ($\varepsilon(T) = C/[(\hbar\omega_0/2k)\coth{(\hbar\omega_0/2kT)} - T_{\textrm{C}}]$~\cite{barrett1952}), this $\varepsilon_B(T)$ function is mainly useful at low enough temperatures, which is not our case.
Moreover, the very approach of a soft optical branch can hardly be used for a description of the order-disorder PT observed in TEACCB-3.

Subtracting the dielectric background, we  found the critical part $\varepsilon_{\text{cr}}(T)$ for TEACCB-3.
The attempts to interpret the $\varepsilon_{\text{cr}}(T)$ data using the Curie-Weiss formula~(\ref{eq:eps-gen-cw}) with either $\gamma = 1$ or $\gamma \neq 1$ have failed: the standard deviation values for both the paraelectric and ferroelectric phases are intolerably high and, moreover, a correlated temperature behavior of the residuals is observed.
Fitting with the alternative formula~(\ref{eq:eps-dif-pt}) is illustrated in figure~\ref{fig:eps-TEACCB} (insert).
Notice that although this semi-empirical formula was derived for the region $T > T_{\textrm{C}}$ only, we  checked its applicability at $T < T_{\textrm{C}}$ for the PT which is close to the second-order.
Then, the term $(T - T_{\textrm{C}})$ in formula~(\ref{eq:eps-dif-pt}) must be changed to $|T - T_{\textrm{C}}|$.
With the similar standard deviation (0.09 and 0.08) and $R^2$ values (0.990 and 0.992) for the regions $T < T_{\textrm{C}}$ and $T > T_{\textrm{C}}$, our fits give the $\gamma$ indices different by only 17\% (1.69 and 1.98, respectively).

However, since a domain-wall contribution to $\varepsilon(T)$ cannot be excluded for the ferroelectric phase originated from the order-disorder PT type~\cite{nakamura1992}, any physical conclusions for that phase must be taken with some precaution.
In this respect, the $\gamma$ parameter derived for the paraelectric phase can be qualified as more reliable.
It is evident that the dielectric anomaly in the solid solution TEACCB-3 is broad and, issuing from the $\gamma$ value, it can be supposed to be close to the relaxor type.
This should imply partial destroying of long-range correlations due to disordered Cl and Br ions in the crystal structure.
It would be interesting to verify our hypothesis of relaxor state with probing frequency dependences $T_{\textrm{C}}(\omega)$ and $\varepsilon_\text{m}(\omega)$.
Finally, according to the more reliable paraelectric data, we  found the `broadening' of the $\varepsilon_{\text{cr}}(T)$ curve: $\delta \approx 5.6$~K$^{\gamma/2}$ or $\delta \approx 5.6$~K at $\gamma \approx 2$ (cf., e. g., with the estimations~\cite{morrisonsinclairwest1999, kovalalemanybriancinbrunckova2003, correakumarkatiyar2007, pilgrimsutherlandwinzer1990}).
Hence, the main features of the $\varepsilon(T)$ function for multiferroic TEACCB-3 are a strong nonlinear background and a considerable diffuseness of the PT, which can be attributed to the structural disorder in this solid solution.

\begin{figure}[!t]
    \centering
    \includegraphics[width=0.6\textwidth]{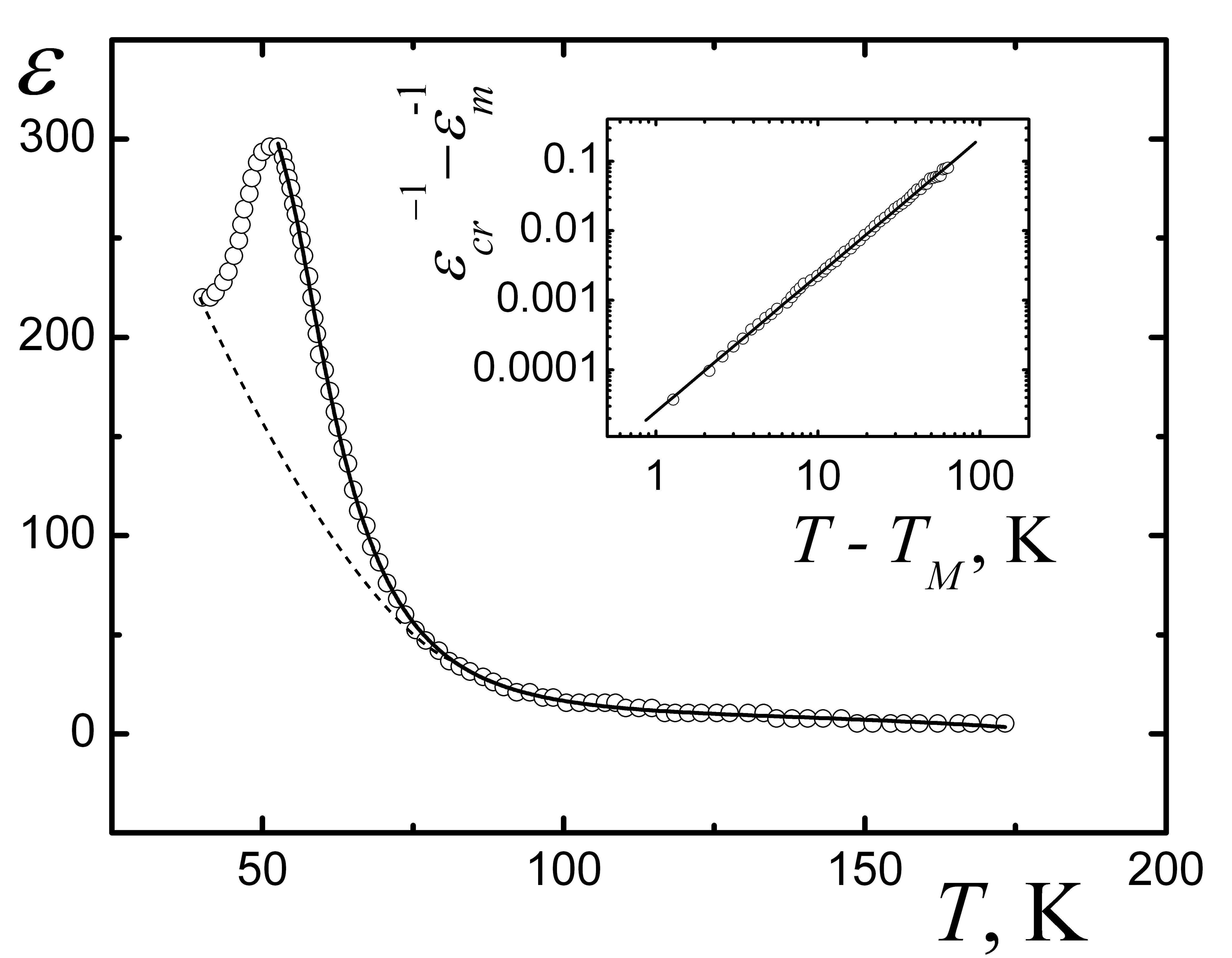}
    \caption{Dependence $\varepsilon(T)$ for the direction of $a$-axis in the Sr$_2$IrO$_4$ crystals, as taken from figure~3b in the work~\cite{chikaraetal2009} ($f = 100$~kHz, magnetic field 0.1~T applied along the $a$-axis).
    Data points correspond to original experiment and smooth curve corresponds to our nonlinear fitting with formula~(\ref{eq:eps-dif-pt}) for the high-temperature region $T > T_M$.
    Insert shows a log-log dependence $1/\varepsilon_{\text{cr}} - 1/\varepsilon_\text{m}$ vs. $(T - T_M)$ at $T > T_M$.}
    \label{fig:eps-SIO}
\end{figure}

The $a$-axis dielectric data for the magnetoelectric Sr$_2$IrO$_4$ single crystals above the temperature $T = T_M$ \cite{chikaraetal2009} (see figure~\ref{fig:eps-SIO}) further illustrate that a proper consideration of a nonlinear background can become important.
Similarly to the data treatment used for the TEACCB-3 crystals, here we  also took advantage of detailed dielectric data outside the PT region.
Namely, we  found the $\varepsilon_B (T)$ function and then fitted the critical part $\varepsilon_{\text{cr}}(T)$, using a linear fitting on a double logarithmic scale (see insert in figure~\ref{fig:eps-SIO}).
It turns out that introduction of a correction $\varepsilon_B(T)$ quadratic in temperature in formula~(\ref{eq:eps-dif-pt}) results in twice as less standard deviation,  compared with the case of no dielectric background.
What is more substantial, a known statistical Wald--Wolfowitz runs test then shows that the residuals become almost one standard deviation closer to their normal (i.e., random) distribution, a condition which must be met if the theoretical model pretends to fit the data.
The excellence of formula~(\ref{eq:eps-dif-pt}) is illustrated by the linear fit in figure~\ref{fig:eps-SIO} (insert) which yeilds the goodness-of-fit parameter $R^2 = 0.9995$.
Finally, we  obtained the critical index $\gamma \approx 2.02$, whereas the breadth of the dielectric anomaly at $T_M$ is $\delta \approx 9.1$~K.

\section{Conclusions} \label{sec:conclusions}

Summarizing, we have experimentally studied  the temperature dependences of the dielectric permittivity for the two single-crystalline compounds, the nonstoichiometric ferroic PGO crystals and the magnetic multiferroic TEACCB-3 crystals, for which the background contributions to the $\varepsilon(T)$ functions are obvious.
General theoretical relations are suggested that consider corrections due to a combined effect of the nonlinear temperature-dependent background $\varepsilon_B(T)$, deviations from the simplest Curie-Weiss law due to structural defects, and a diffuseness of PT.
The importance of simultaneously taking account of all those factors is demonstrated by successful statistically grounded quantitative interpretation of the experimental 
$\varepsilon(T)$ dependences for both materials under test, with reasonable values obtained for the fitting parameters which correlate satisfactorily for both structural phases.
Our data confirm that a constant dielectric-background approximation can turn out to be insufficient.
In particular, this is proved by a comparison with our earlier data for PGO 
\cite{elit2018}.
It is important that consideration of the temperature-dependent $\varepsilon_B(T)$ term for the magnetoelectric Sr$_2$IrO$_4$ single crystals notably improves the accuracy obtained for the PT parameters.
Summing up, the $\varepsilon(T)$ dependence for PGO turns out to be more or less typical of the ferroelectric PTs, whereas TEACCB-3 reveals some features of the diffuse relaxor-type behavior.

The other conclusion drawn from our results is of a general methodological character: even though the background dielectric terms can be small enough with respect to the critical PT-related dielectric term, a complete neglect of the $\varepsilon_B$ term (or sometimes even a neglect of the temperature-dependent $\varepsilon_B(T)$ term) is hardly desirable.
A good illustration of relative importance of the background in the dielectric permittivity is given in table~\ref{tab:eps-comparison}.
The absolute (or `apparent') maxima $\varepsilon_{\text{max}}$ of the $\varepsilon(T)$ dependences detected in the experiments are compared with the maxima $\varepsilon_\text{m}$ associated with the `true' (i. e., background-free) critical behavior $\varepsilon_{\text{cr}}(T)$, which have been found from our fittings.
The difference of those parameters is due to the background terms.
As seen from table~\ref{tab:eps-comparison}, the dielectric background in PGO dominates and the term $\varepsilon_\text{m}$ amounts to only 15\% of $\varepsilon_{\text{max}}$.

\begin{table}[htb]
    \caption{Absolute dielectric maxima $\varepsilon_{\text{max}}$ detected in the experiments and the corresponding maxima $\varepsilon_\text{m}$ associated solely with the critical behavior.}
    \label{tab:eps-comparison}
    \vspace{3mm}
    \centering
    \begin{tabular}{|c|c|c|}
    \hline
    Crystal & $\varepsilon_{\text{max}}$ & $\varepsilon_\text{m}$ \\
    \hline
    PGO & 50.9 & 7.8 * \\
    TEACCB-3 & 65.2 & 45.3 \\
    Sr$_2$IrO$_4$ & 297.0 & 145.5 \\
    \hline
    \end{tabular}
    \begin{minipage}[c]{0.8\textwidth}
    	 \vspace{3mm}
    \footnotesize{* Note that this value has been obtained from a less accurate fitting with formula~(\ref{eq:eps-dif-pt}), which is rejected statistically.}
    \end{minipage}
\end{table}

Therefore, consideration of the PT-independent contributions can greatly improve the quantitative interpretation of the experimental data and can yield a higher accuracy for the PT parameters, including the critical index $\gamma$.
Of course, any serious theoretical analysis of the heat capacity of ferroics cannot be done without taking  account of the background lattice contributions.
We would argue that the same is advisable for the case of dielectric permittivity, at least for such materials as weakly polar, finite-sized (or confined), nonstoichiometric and improper ferroics, ferroics with a noticeable defect concentrations, and multiferroics.

\bibliography{references}
\bibliographystyle{cmpj}

\newpage
\ukrainianpart

\title{Поправки на нелінійний фон до діелектричної проникності фероїків та мультифероїків}
\author{І. С. Гірник\refaddr{label1}, Б. І. Горон\refaddr{label1,label2}, В. Б. Капустяник\refaddr{label1}, О. С. Кушнір\refaddr{label2}, Р. Ю. Шопа\refaddr{label4}} 
\addresses{
	\addr{label1} Фізичний факультет, Львівський національний університет імені Івана Франка, вул. Драгоманова, 50, 79005 Львів, Україна
	\addr{label2} Факультет електроніки та комп'ютерних технологій, Львівський національний університет імені Івана Франка, вул. ген. Тарнавського, 107, 79017 Львів, Україна
	\addr{label4} Відділ складних систем, Національний центр ядерних досліджень, 05--400 Отвоцьк-Швієрк, Польща}

\makeukrtitle

\begin{abstract}
	Проведено температурні вимірювання діелектричної проникності для нестехіометричного сегнетоелектрика германату свинцю Pb$_{4.95}$Ge$_3$O$_{11}$ і твердого розчину мультифероїка [N(C$_2$H$_5$)$_4$]$_2$CoClBr$_3$.
	На відміну від даних теплоємності, аналіз діелектричної проникності зазвичай проводять, виходячи з припущення про те, що діелектричний `фон' нехтовно малий, порівняно з критичною складовою.
	У цій роботі кількісно проінтерпретовано діелектричні властивості згаданих вище кристалів, а також відповідні дані літератури для кристалів мультифероїка Sr$_2$IrO$_4$, використовуючи узагальнені формули Кюрі-Вейса, в яких поєднані поправки на залежний від температури нелінійний діелектричний фон, модифікований критичний індекс електричної сприйнятності та розмитий характер фазового переходу.
	Ми стверджуємо, що врахування залежного від температури діелектричного фону може значно вдосконалити кількісний аналіз фазових переходів для низки класів фероїків.
	\keywords сегнетоелектрики, фазові переходи, розмиті фазові переходи, діелектрична проникність, критична поведінка, германат свинцю
\end{abstract}

\end{document}